\documentclass{article}
\usepackage{cite}
\usepackage{graphicx}
\usepackage{dcolumn}

\begin{document}

\title{Comment on: \textquotedblleft Tunneling of coupled methyl quantum
rotors in 4-methylpyridine: Single rotor potential versus coupling
interaction''. J. Chem. Phys. \textbf{147}, 194303 (2017)}
\author{Paolo Amore\thanks{%
e--mail: paolo@ucol.mx} \\
Facultad de Ciencias, CUICBAS, Universidad de Colima,\\
Bernal D\'{\i}az del Castillo 340, Colima, Colima,Mexico \\
and \\
Francisco M. Fern\'andez\thanks{%
e--mail: fernande@quimica.unlp.edu.ar} \\
INIFTA, Divisi\'{o}n Qu\'{\i}mica Te\'{o}rica,\\
Blvd. 113 y 64 (S/N), Sucursal 4, Casilla de Correo 16,\\
1900 La Plata, Argentina}
\maketitle

\begin{abstract}
We analyze the results obtained recently by Khazael and Sebastiani (J. Chem.
Phys. \textbf{147}, 194303 (2017)) for two coupled methyl rotors. We show
that one of the avoided crossings obtained by those authors is in fact a
true crossing between states of different symmetry. Present analysis is
supposed to be clearer because it is based on an independent calculation of
eigenvalues for each irreducible representation. The application of group
theory also enables a more accurate calculation by means of smaller matrix
representations of the Hamiltonian operator.
\end{abstract}

\section{Introduction}

\label{sec:intro}

In a recent paper Khazaei and Sebastiani\cite{KS17} (KS from now on) studied
the influence of rotational coupling between a pair of methyl rotors on the
tunneling spectrum in condensed phase. They considered the simplest model in
which two hindered $C_{3}$ rotors interact by means of a periodic potential.
They calculated some of the eigenvalues of the model by means of the
Rayleigh-Ritz variational method based on a Fourier basis set. The authors
analyzed the lowest energy levels in different regions of coupling strength
and found two avoided crossings between two pairs of them. According to KS
the avoided crossing closest to the origin is due to a pair of energy levels
of symmetry $AE$ and $EE$. However, states of different symmetry are
expected to undergo actual crossings and not avoided ones. Avoided crossings
take place between states of the same symmetry\cite{T37,RNBB72,F14} (and
references therein). This confusion probably arose from the fact that KS did
not calculate the energy levels for each irreducible representation (irrep)
separately but used a Fourier basis set that contains all of them. Such a
calculation is unnecessarily demanding because it leads to a much larger
matrix representation of the Hamiltonian operator. In addition to it, it is
not so easy to distinguish a true crossing from an avoided one in a
calculation that includes all the irreps. If, on the other hand, one carries
out a calculation for every irrep, then one immediately realizes that any
pair of energy levels that approach each other will undergo an avoided
crossing no matter how close they approach each other.

The purpose of this comment is to carry out a Rayleigh-Ritz variational
calculation with a matrix representation of the Hamiltonian operator for
every irrep. In section~\ref{sec:model} we outline the coupled-rotors model
and briefly discuss the symmetry of its states. In section~\ref{sec:results}
we compare present results with those of KS and draw conclusions.

\section{The model and its symmetry}

\label{sec:model}

The Hamiltonian for two interacting rotors studied by KS is
\begin{eqnarray}
H &=&T_{1}+T_{2}+V_{1}(\phi _{1})+V_{2}(\phi _{2})+V_{12}(\phi _{1},\phi
_{2}),  \nonumber \\
T_{i} &=&-B\frac{\partial ^{2}}{\partial \phi _{i}},\;V_{i}(\phi
_{i})=V_{3}\cos (3\phi _{i}),\;i=1,2,  \nonumber \\
V_{12}(\phi _{1},\phi _{2}) &=&W_{3}\cos (3\phi _{1}-3\phi _{2}),
\label{eq:H_2D}
\end{eqnarray}%
where $B=\hbar ^{2}/(2I)$ is the rotational constant, $I$ is a suitable
moment of inertia for an individual rotor, $V_{3}$ is a measure of the
hindering potential for each rotor and $W_{3}$ is the strength of the
interaction\cite{KS17} (and references therein).

The Hamiltonian operator $H$ is invariant under the unitary transformations $%
U_{i}^{-1}\phi _{i}U_{i}=\phi _{i}+2\pi /3$, $i=1,2$; therefore the symmetry
group is at least $C_{3}\otimes C_{3}$. Since the symmetry species for $C_{3}
$ are $A$ and $E$ (which may be separated into $E_{a}$ and $E_{b}$ because $E
$ is two-fold degenerate)\cite{C90} then the eigenfunctions of $H$ can be
classified as
\begin{equation}
\begin{array}{lrr}
& s_{1} & s_{2} \\
AA & 0 & 0 \\
AE_{a} & 0 & 1 \\
E_{a}A & 1 & 0 \\
AE_{b} & 0 & -1 \\
E_{b}A & -1 & 0 \\
E_{a}E_{b} & 1 & -1 \\
E_{b}E_{a} & -1 & 1 \\
E_{a}E_{a} & 1 & 1 \\
E_{b}E_{b} & -1 & -1%
\end{array}%
,  \label{eq:symmetry_pair}
\end{equation}%
where $s_{1}$ and $s_{2}$ are symmetry quantum numbers that enable us to
expand every eigenfunction $\psi (\phi _{1},\phi _{2})$ in a Fourier series
of the form
\begin{eqnarray}
\psi _{n_{1},n_{2},s_{1},s_{2}}(\phi _{1},\phi _{2}) &=&\sum_{j_{1}=-\infty
}^{\infty }\sum_{j_{2}=-\infty }^{\infty
}c_{j_{1},j_{2},n_{1},n_{2},s_{1},s_{2}}\varphi _{j_{1},s_{1}}(\phi
_{1})\varphi _{j_{2},s_{2}}(\phi _{2}),  \nonumber \\
\varphi _{j,s}(\phi ) &=&\frac{1}{\sqrt{2\pi }}\exp \left[ \left(
3j+s\right) i\phi \right] ,\;s=0,\pm 1.  \label{eq:eigenfunctions}
\end{eqnarray}%
Notice that we can separate the matrix representation of the Hamiltonian
operator in $9$ different and wholly independent cases which makes the
calculation considerably more efficient. In this way, we obtain more
accurate results with matrices of smaller dimension and the analysis\ of the
spectrum is remarkably simpler as shown below. On the other hand, KS
resorted to the basis set $B^{KS}=\left\{ \exp \left[ i\left( m\phi
_{1}+n\phi _{2}\right) \right] /(2\pi ),\;m,n=0,\pm 1,\ldots \right\} $
which may lead to some confusion in the analysis of the spectrum.

Two other symmetry operations, parity $P_{a}:(\phi _{1},\phi
_{2})\rightarrow (-\phi _{1},-\phi _{2})$ and permutation $P_{e}:(\phi
_{1},\phi _{2})\rightarrow (\phi _{2},\phi _{1})$, also leave the
Hamiltonian invariant $PHP^{-1}=H$, where $P^{-1}$ stands for the inverse of
$P$. Notice that for these operators it also follows that $P^{-1}=P^{\dagger
}=P$ and that $[H,P]=0$. Therefore, if $\psi $ is an eigenfunction of $H$
with eigenvalue $\epsilon $ then $P\psi $ is also an eigenfunction of $H$
with the same eigenvalue as shown by $PH\psi =PHP^{-1}P\psi =HP\psi
=\epsilon P\psi $. It is clear that if $\psi $ and $P\psi $ are linearly
independent, then they are degenerate solutions of the Schr\"{o}dinger
equation. If we take into account the effect of these operations on the
symmetry of the states
\begin{equation}
\begin{array}{lrr}
& P_{a} & P_{e} \\
AA & AA & AA \\
AE_{a} & AE_{b} & E_{a}A \\
E_{a}A & E_{b}A & AE_{a} \\
AE_{b} & AE_{a} & E_{b}A \\
E_{b}A & E_{a}A & AE_{b} \\
E_{a}E_{b} & E_{b}E_{a} & E_{b}E_{a} \\
E_{b}E_{a} & E_{a}E_{b} & E_{a}E_{b} \\
E_{a}E_{a} & E_{b}E_{b} & E_{a}E_{a} \\
E_{b}E_{b} & E_{a}E_{a} & E_{b}E_{b}%
\end{array}%
,  \label{eq:symmetry_transformations}
\end{equation}%
we conclude that each state $AA$ is nondegenerate, $\left\{
AE_{a},AE_{b},E_{a}A,E_{b}A\right\} $ are four-fold degenerate, $\left\{
E_{a}E_{b},E_{b}E_{a}\right\} $ are two-fold degenerate as well as $\left\{
E_{a}E_{a},E_{b}E_{b}\right\} $ in agreement with the statement of H\"{a}%
usler and H\"{u}ller\cite{HH85}. For brevity, from now on we refer to these
states simply as $AA$, $AE$, $EE^{\prime }$ and $EE$, respectively. When $%
W_{3}=0$ the two rotors are uncoupled and the Hamiltonian operator $%
H=H_{1}+H_{2}$ is a sum of the Hamiltonian operators $H_{i}$ for the
individual rotors. The energy levels are sums of eigenvalues $\epsilon
_{k}^{(i)}$ of each $H_{i}$, $i=1,2$, so that we may have combinations like $%
\epsilon _{mn}(E_{X}E_{Y})=\epsilon _{m}^{(1)}(E_{X})+\epsilon
_{n}^{(2)}(E_{Y})$. Each of these levels is four-fold degenerate because $%
E_{X}E_{Y}$ can be $E_{a}E_{a},E_{a}E_{b},E_{b}E_{a},E_{b}E_{b}$ and,
according to what was said above, the interaction partially splits it into
two two-fold degenerate levels (namely $EE$ and $EE^{\prime }$). Curiously,
KS state that this rigorous analysis based on group theory is a \textit{%
speculation} confirmed by their numerical calculations.

The approximate finite matrix representation of the Hamiltonian can be
written in terms of sums of Kronecker products
\begin{eqnarray}
\mathbf{H} &=&\mathbf{T}_{1}\otimes \mathbf{I}+\mathbf{I\otimes T}%
_{2}+V_{3}\left( \mathbf{C}_{1}\otimes \mathbf{I}+\mathbf{I\otimes C}%
_{2}\right)  \nonumber \\
&&+W_{3}\left( \mathbf{C}_{1}\otimes \mathbf{C}_{2}+\mathbf{S}_{1}\mathbf{%
\otimes S}_{2}\right) ,  \label{eq:H_matrix}
\end{eqnarray}%
where $\mathbf{T}$, $\mathbf{C}$, $\mathbf{S}$ are $(2N+1)\times (2N+1)$
matrix representations of the operators $-B\partial ^{2}/\partial \phi ^{2}$%
, $\cos (\phi )$ and $\sin (\phi )$, respectively, in the basis set $\left\{
\varphi _{j,s}(\phi ),\;j=-N,-N+1,\ldots ,N\right\} $ and $\mathbf{I}$ is
the identity matrix of the same dimension. Consequently, the dimension of
the matrix representation $\mathbf{H}$ is $(2N+1)^{2}\times (2N+1)^{2}$ and
the pair of symmetry quantum numbers $s_{1}$ and $s_{2}$ for the left and
right factors in each product determines the symmetry of the state as shown
in equation (\ref{eq:symmetry_pair}).

\section{Results and conclusions}

\label{sec:results}

In order to compare present results with those of KS we calculate the energy
differences $\Delta \epsilon _{j}(XY)=\epsilon _{j}(XY)-\epsilon _{0}(AA)$, $%
j=0,1,\ldots $, where $X,Y=A,E_{a},E_{b}$. Besides, we choose $B=0.654\,meV$%
, $V_{3}=5\,meV$ and a wide range of values of $W_{3}\geq 0$ in agreement
with the calculation of KS.

\begin{figure}[tbp]
\begin{center}
\includegraphics[width=6cm]{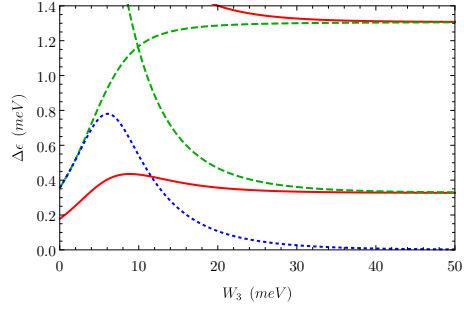}
\end{center}
\caption{Lowest $\Delta\protect\epsilon$ for the symmetries $AE$ (solid
line, red), $EE^{\prime }$ (dashed line, blue) and $EE$ (dashed line, green)}
\label{fig:eks}
\end{figure}

\begin{figure}[tbp]
\begin{center}
\includegraphics[width=6cm]{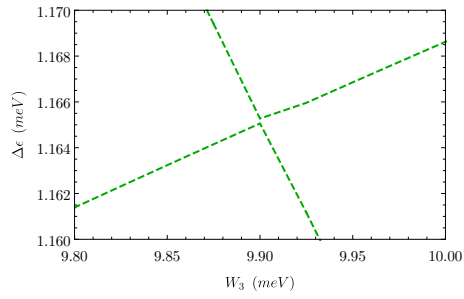}
\end{center}
\caption{Avoided crossing closest to the origin}
\label{fig:ac}
\end{figure}

Figure~\ref{fig:eks} shows values of $\Delta \epsilon _{j}$ for the first $15
$ states and is equivalent to Figure~3 of KS. $\Delta \epsilon _{0}(AA)$ is
nondegenerate and, by definition, identical to zero for all $W_{3}$. The
four-fold degenerate $\Delta \epsilon _{0}\left( AE\right) $ exhibits a
maximum. The set of four states $\{EE^{\prime },EE\}$ splits when $W_{3}>0$
in such a way that $\Delta \epsilon _{0}\left( EE\right) $ increases and $%
\Delta \epsilon _{0}\left( EE^{\prime }\right) $ reaches a maximum after
which it decreases with $W_{3}$. The latter crosses $\Delta \epsilon
_{0}\left( AE\right) $ at $W_{3}=W_{3}^{c}$ so that $\Delta \epsilon
_{0}\left( AE\right) <$ $\Delta \epsilon _{0}\left( EE^{\prime }\right) $
when $W_{3}<W_{3}^{c}$ and $\Delta \epsilon _{0}\left( AE\right) >\Delta
\epsilon _{0}\left( EE^{\prime }\right) $ otherwise. This is a true crossing
and not an avoided one as stated by KS. The two-fold degenerate levels $%
\Delta \epsilon _{0}\left( EE^{\prime }\right) $ and $\Delta \epsilon
_{1}\left( EE^{\prime }\right) $ approach each other quite closely and
exhibit and avoided crossing but always $\Delta \epsilon _{0}\left(
EE^{\prime }\right) <\Delta \epsilon _{1}\left( EE^{\prime }\right) $ (they
have a maximum and a minimum, respectively, at the avoided crossing). It can
be proved that they coalesce at an exceptional point $W_{3}^{EP}$ in the
complex $W_{3}$ plane\cite{HS90}. This is the avoided crossing closest to
the origin and the only one in Figure~3 of KS. This avoided crossing is
magnified in Figure~\ref{fig:ac}. In Figure~\ref{fig:eks} we also appreciate
the four-fold degenerate level $\Delta \epsilon _{1}\left( AE\right) $ (the
highest one shown by that energy scale).

\begin{figure}[tbp]
\begin{center}
\includegraphics[width=6cm]{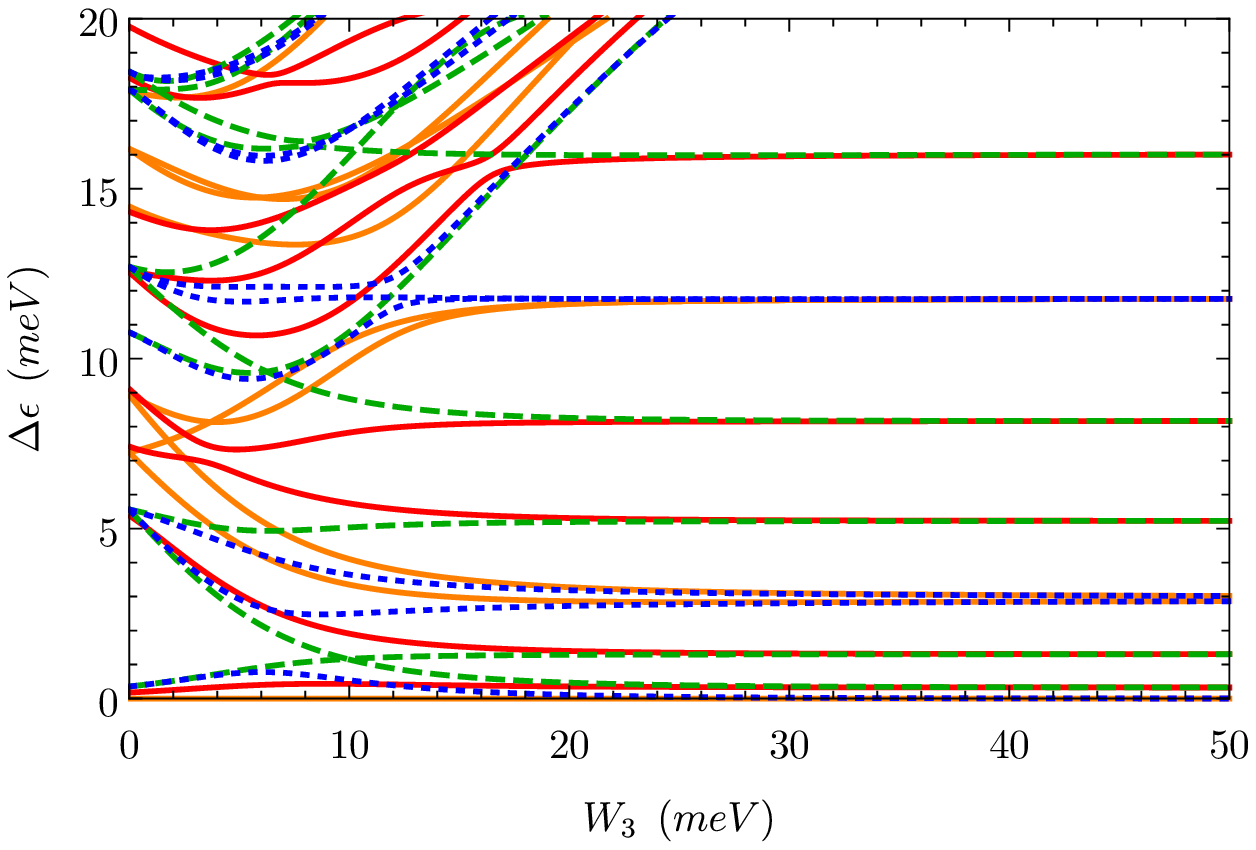}
\end{center}
\caption{Lowest $\Delta\protect\epsilon$ for the symmetries $AA$ (solid
line, orange), $AE$ (solid line, red) $EE^{\prime }$ (dashed line, blue) and
$EE$ (dashed line, green)}
\label{fig:e}
\end{figure}

Figure~\ref{fig:e} shows some more energy levels that exhibit crossings
(different symmetry) and avoided crossings (same symmetry). Because of the
former the order of several energy levels changes with $W_{3}$ and for this
reason we prefer to label the members of each irrep separately so that $%
\Delta \epsilon _{j}\left( XY\right) <\Delta \epsilon _{j+1}\left( XY\right)
$ for all $W_{3}$.

As stated in the introduction, the purpose of this comment is the analysis
of the energy levels of the simple coupled-rotors model (\ref{eq:H_2D}) in a
clearer way than that provided by KS. To this end we resorted to a separate
calculation of the energy levels for each irrep because it is difficult to
distinguish a true crossing from an avoided one if one uses the Fourier
basis set $B^{KS}$ chosen by KS. The reason lies in the difficulty of
tracking a pair of energy levels through a crossing or when they approach
each other too much. It is not surprising that KS confused a true crossing
for an avoided one when they based their conclusions only on numerical
calculations with the basis $B^{KS}$. Our figures clearly display crossings
and avoided crossings because it is possible to use different kinds of lines
or symbols for different irreps when one treats them separately.

Another relevant point of this comment is the analysis of the spectrum of
the model in terms of group theory that is a useful tool for the prediction
of degeneracies and their splitting when a Hamiltonian operator is perturbed
by a known interaction. Without any calculation one knows that states of
different symmetry are not expected to give rise to avoided crossings\cite%
{T37,RNBB72,F14}. In fact, different irreps behave as if they were
completely independent problems because they do not exhibit any interaction.
They can, therefore, be treated separately reducing the dimension of the
matrix representation $\mathbf{H}$ that increases with $N$ as $(2N+1)^{2}$.
In the present case we roughly need $N/9$ for each irrep to achieve the same
accuracy.

\section*{Acknowledgements}
The research of P.A. was supported by Sistema Nacional de Investigadores (M\'exico).

\end{document}